\documentclass{PoS}

\usepackage{amssymb,pslatex}
\usepackage{amsmath,marvosym}
\usepackage{braket}
\renewcommand{\d}{\ensuremath{\mathrm{d}}}
\newcommand{\p}{\partial}
\newcommand{\GZ}{\ensuremath{\mathrm{GZ}}}
\newcommand{\gf}{\ensuremath{\mathrm{gf}}}
\newcommand{\YM}{\ensuremath{\mathrm{YM}}}

\title{Evidence of BRST-Symmetry Breaking in Lattice Minimal Landau Gauge}

\ShortTitle{BRST-Symmetry Breaking}

\author{\speaker{Attilio Cucchieri}\\
        Instituto de F\'\i sica de S\~ao Carlos, Universidade de S\~ao Paulo,
        Caixa Postal 369, 13560-970 S\~ao Carlos, SP, Brazil\\
        E-mail: \email{attilio@ifsc.usp.br}}

\author{David Dudal\\
        KU Leuven Campus Kortrijk - KULAK, Department of Physics, Etienne Sabbelaan 53, 8500 Kortrijk, Belgium\\
        Ghent University, Department of Physics and Astronomy,
        Krijgslaan 281-S9, 9000 Gent, Belgium\\  
        E-mail: \email{david.dudal@ugent.be}}

\author{Tereza Mendes\\
        Instituto de F\'\i sica de S\~ao Carlos, Universidade de S\~ao Paulo,
        Caixa Postal 369, 13560-970 S\~ao Carlos, SP, Brazil\\
        E-mail: \email{mendes@ifsc.usp.br}}

\author{Nele Vandersickel\\
        Ghent University, Department of Physics and Astronomy,
        Krijgslaan 281-S9, 9000 Gent, Belgium\\
        E-mail: \email{nele.vandersickel@ugent.be}}

\abstract{By evaluating the so-called Bose-ghost propagator, we present the
first numerical evidence of BRST-symmetry breaking for Yang-Mills theory
in minimal Landau gauge, i.e.\ due to the restriction of the functional integration
to the first Gribov region in the Gribov-Zwanziger approach. Our data
\cite{Cucchieri:2014via} are well described by a simple fitting function,
which can be related to a massive gluon propagator in combination with an
infrared-free (Faddeev-Popov) ghost propagator. As a consequence,
the Bose-ghost propagator, which has been proposed as a carrier of
the confining force in minimal Landau gauge, displays a $1/p^4$ singularity
in the infrared limit.}

\FullConference{The 32nd International Symposium on Lattice Field Theory,\\
		23-28 June, 2014\\
		Columbia University New York, NY}

\begin{document}


\section{Introduction}

The search for confinement signatures in the infrared (IR) behavior
of Green's functions is a longstanding {\em holy grail} of Yang-Mills theories
and Quantum Chromodynamics. Indeed, as stressed by West \cite{West:1982bt},
already the area-law criterion of Wilson \cite{Wilson:1974sk} was
partially based on the presumed IR behavior of the gluon propagator.
This viewpoint was also addressed by several studies, based on functional
approaches, such as the works of Mandelstam \cite{Mandelstam:1979xd}, Baker,
Ball and Zachariasen \cite{Baker:1980gf,Baker:1980gd}, Cornwall
\cite{Cornwall:1981zr}, Stingl \cite{Stingl:1985hx}, to name a few.

Among the possible gauge-fixing conditions considered for the evaluation
of Green's functions in Yang-Mills theories, the so-called minimal Landau gauge has
attracted a great deal of attention since Gribov's work \cite{Gribov:1977wm}.
Let us recall that, in this case, the gauge condition is implemented
by restricting the functional integral over gauge-field configurations
to the so-called first Gribov region $\Omega$, i.e.\ to the set of transverse
configurations for which the Faddeev-Popov matrix ${\cal M}^{ab}(x,y)$
is non-negative. In a numerical simulation
(on the lattice) this can be easily obtained by minimizing a suitable
functional (see for example Ref.\ \cite{Giusti:2001xf}).
On the other hand, in analytic studies, this restriction is achieved
by adding a nonlocal term $S_{\mathrm{h}}$, the horizon function, to the
usual Landau gauge-fixed Yang-Mills action $S_{\YM} + S_{\gf}$. One thus
obtains the Gribov-Zwanziger (GZ) action $S_{\GZ} = S_{\YM} + S_{\gf}
+ \gamma^4 S_{\mathrm{h}}$, where the massive parameter $\gamma$, known as the
Gribov parameter, is dynamically determined by a self-consistent condition,
the horizon condition.

In order to localize the GZ action \cite{Vandersickel:2012tz} one introduces a pair
of complex-conjugate bosonic fields $(\overline{\phi}^{ac}_{\mu},
\phi^{ac}_{\mu} )$ and a pair of Grassmann complex-conjugate fields
$(\overline{\omega}^{ac}_{\mu}, \omega^{ac}_{\mu} )$. Then, the GZ action
can be written as $ S_{\GZ} = S_{\YM} + S_{\gf} + S_{\mathrm{aux}}
+ S_{\gamma}$, where
\begin{eqnarray}
\label{saux}
\!\!\!\!\!\!\!\!\!\! S_{\mathrm{aux}} & = & \int \d^{\rm 4} x \,
\Bigl[ \overline{\phi}_{\mu}^{ac} \, \p_{\nu} \left( D_{\nu}^{ab}
                     \phi^{bc}_{\mu} \right) -
     \overline{\omega}_{\mu}^{ac} \, \p_{\nu} \left( D_{\nu}^{ab}
                     \omega^{bc}_{\mu} \right) 
                   - g_0 \left( \p_{\nu}
         \overline{\omega}_{\mu}^{ac} \right) f^{abd} \,
                D_\nu^{be} \eta^e \phi_{\mu}^{dc} \Bigr] \\[1mm]
\!\!\!\!\!\!\!\!\!\! S_{\gamma} &
                     = &
                        \int \d^{\rm 4}x \,
           \Bigl[ \gamma^{2}
                 D^{ba}_{\nu} \Bigl( \phi_{\nu}^{ab}
           + \overline{\phi}_{\nu}^{ab} \Bigr)
        - 4 \left( N_c^{2} - 1 \right) \gamma^{4} \Bigr] \, .
\label{eq:Sofgamma}
\end{eqnarray}

Under the nilpotent BRST variation $s$, the four
auxiliary fields form two BRST doublets, i.e.\
\begin{equation}
s \, \phi^{ac}_{\mu} \, = \, \omega^{ac}_{\mu} \, ,
\qquad \qquad \qquad s \, \omega^{ac}_{\mu} \, = \, 0
\end{equation}
and
\begin{equation}
s \, \overline{\omega}^{ac}_{\mu} \, = \, \overline{\phi}^{ac}_{\mu} \, ,
\qquad \qquad \qquad s \, \overline{\phi}^{ac}_{\mu} \, = \, 0 \, ,
\end{equation}
giving rise to a BRST quartet. At the same time, one can check that
the localized GZ theory is not BRST-invariant. Indeed, while
$s \, (S_{\YM} + S_{\gf} + S_{\mathrm{aux}}) = 0$, one finds that
$s \, S_{\gamma} \propto \gamma^2 \neq 0$. Since
a nonzero value for the Gribov parameter $\gamma$ is implied by the
restriction of the functional integration to the Gribov region
$\Omega$, it is clear that BRST-symmetry breaking is expected, as a direct
consequence of this restriction.


\section{The Bose-Ghost Propagator}

In order to study numerically the effect of the BRST-breaking
term $S_{\gamma}$, one can consider the expectation value of a
BRST-exact quantity. For example, the correlation function
\begin{equation}
Q^{abcd}_{\mu \nu}(x,y) \, = \,
\braket{ \, s (\, \phi^{ab}_{\mu}(x) \,
\overline{\omega}^{cd}_{\nu}(y)) \, } \, = \,
\braket{ \,
 \omega^{ab}_{\mu}(x) \, \overline{\omega}^{cd}_{\nu}(y) \, + \,
 \phi^{ab}_{\mu}(x) \, \overline{\phi}^{cd}_{\nu}(y) \, }
\end{equation}
should have a zero expectation value for a BRST-invariant theory, but
it does not necessarily vanish if BRST symmetry is broken
\cite{Sorella:2009vt}.  Indeed, one can verify \cite{Vandersickel:2012tz,
Gracey:2010df} that at tree level (and in momentum space)
\vskip 0.5mm
\begin{equation}
\label{eq:Qabcd}
Q^{abcd}_{\mu \nu}(p,p') \, = \,
 \frac{ \left(2 \pi\right)^4 \delta^{(4)}(p + p')
        \, g_0^2 \, \gamma^4 f^{abe} f^{cde} P_{\mu \nu}(p)}{
          p^2 \, \left(p^4 + 2 g_0^2 N_c \gamma^4 \right)} \, ,
\end{equation}
where $P_{\mu \nu}(p)$ is the usual transverse projector. Thus,
this propagator is proportional to the Gribov parameter $\gamma$,
i.e.\ its nonzero value is clearly related to the breaking of the
BRST symmetry in the GZ theory. One should also
recall that this Bose-ghost propagator has been proposed as a carrier
of long-range confining force in minimal Landau gauge
\cite{Zwanziger:2009je,Furui:2009nj,Zwanziger:2010iz}.

\vskip 2.5mm
On the lattice one does not have direct access to the auxiliary
fields $(\overline{\phi}^{ac}_{\mu}, \phi^{ac}_{\mu} )$ and
$(\overline{\omega}^{ac}_{\mu}, \omega^{ac}_{\mu} )$.
On the other hand, by 1) adding suitable sources to the GZ action,
2) explicitly integrating over the four auxiliary fields
and 3) taking the usual
functional derivatives with respect to the sources,\footnote{This
is analogous to the evaluation of the Faddeev-Popov correlation
function, i.e. the ghost propagator, for which, on the lattice, one
uses the relation
\begin{equation}
\braket{ \, \overline{c}^{a}(x) \,
c^{b}(y)) \, } \, = \, \braket{ ( {\cal M}^{-1} )^{ab}(x,y) } \, .
\nonumber
\end{equation}
}
one can verify that \cite{Zwanziger:2009je}
\begin{equation}
\!\!\!\!\!\!\!\!
Q^{abcd}_{\mu \nu}(x-y) \, = \, \gamma^4 \, \left\langle \,
       R^{a b}_{\mu}(x) \, R^{c d}_{\nu}(y) \, \right\rangle \, ,
\label{eq:Qprop}
\end{equation}
where
\begin{equation}
R^{a c}_{\mu}(x) \, = \, \int \d^{\rm 4} z \,
         ( {\cal M}^{-1} )^{ae}(x,z) \, B^{ec}_{\mu}(z)
\end{equation}
and
\begin{equation}
B^{bc}_{\nu}(x) \, = \, g_0 \, f^{b e c} \, A^{e}_{\nu}(x) \, .
\end{equation}


\section{Numerical Simulations}

\begin{figure}
\begin{center}
\vskip -0.6cm
\hskip -5mm
\includegraphics[trim=55 0 40 0, clip, scale=1.00, width=0.7\linewidth]{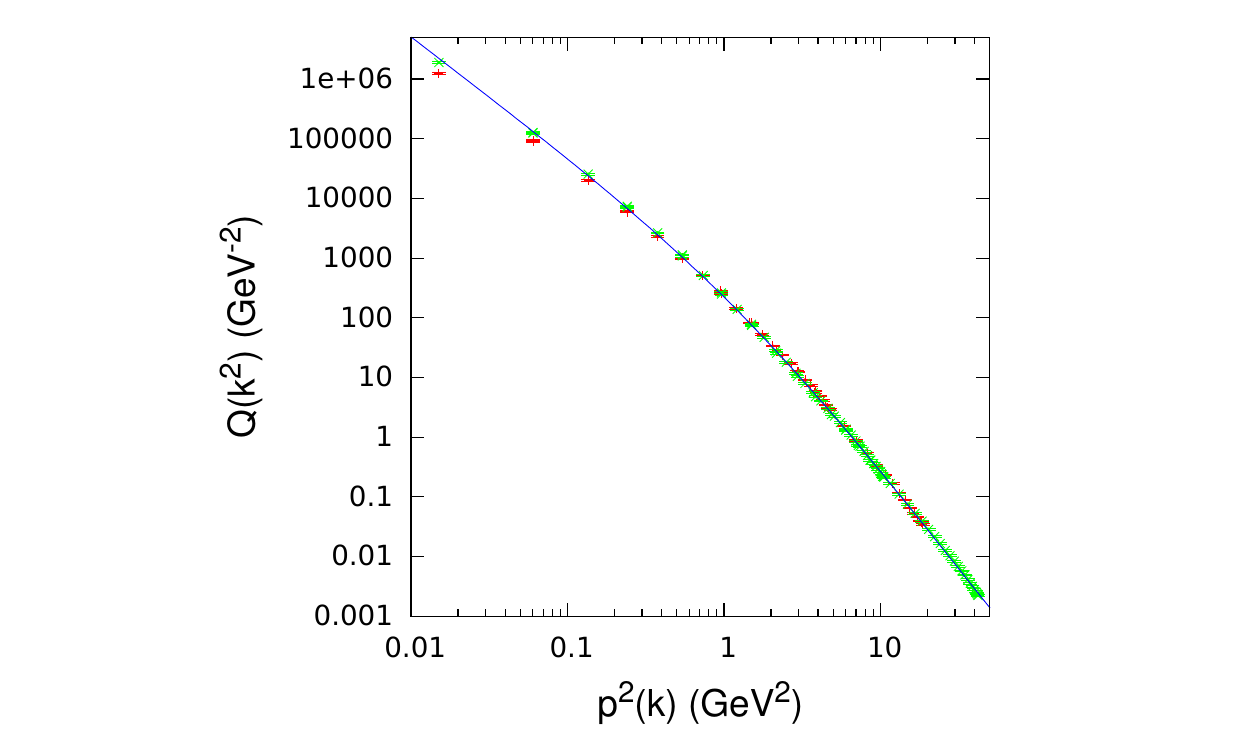}
\vskip 3mm
\caption{
\label{figs1}
Data for $\beta = 2.2$ and $V = 48^4$ ({\bf \textcolor{red}{$+$}}) matched
\cite{Leinweber:1998uu,Cucchieri:2003di} with data for $\beta = 2.34940204$ and $V = 72^4$
({\bf \textcolor{green}{$\times$}}), fitted using Eq.\ (\protect\ref{eq:fit})
with $t = 3.2(0.3) (GeV^2)$, $u = 3.6(0.4) (GeV)$, $s = 46(13) (GeV^2)$ and
$c = 114(13)$.
}
\end{center}
\end{figure}

\begin{figure}
\begin{center}
\vskip -0.6cm
\hskip -5mm
\includegraphics[trim=55 0 40 0, clip, scale=1.00, width=0.7\linewidth]{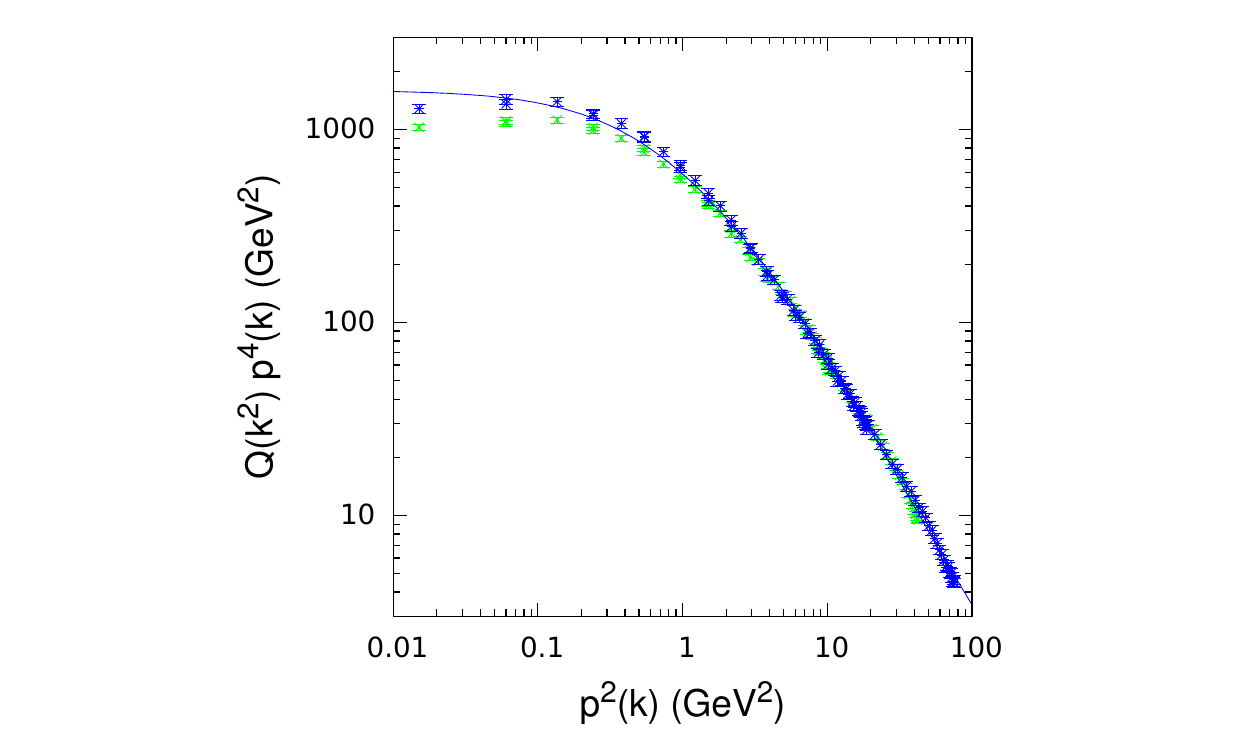}
\vskip 3mm
\caption{
\label{figs2}
Data for $\beta = 2.34940204$ and $V = 72^4$ ({\bf \textcolor{green}{$\times$}})
matched \cite{Leinweber:1998uu,Cucchieri:2003di} with data $\beta = 2.43668228$ and $V = 96^4$
({\bf \textcolor{blue}{$*$}}), fitted using Eq.\ (\protect\ref{eq:fit})
with $t = 3.0(0.2) (GeV^2)$, $u = 3.9(0.3) (GeV)$, $s = 58.0(9.8) (GeV^2)$ and
$c = 247(16)$.
}
\end{center}
\end{figure}

We evaluated \cite{Cucchieri:2014via} the Bose-ghost propagator,
defined in Eq.\ (\ref{eq:Qprop}) above, in momentum space ---modulo the global
factor $\gamma^4$--- using numerical simulations in the SU(2) case. In
order to check discretization and finite-volume effects, we considered
three different values of the lattice coupling $\beta$ and five
different physical volumes, ranging from about $(3.366 \, fm)^4$ to
$(10.097 \, fm)^4$. Numerical results for the scalar function $Q(k^2)$,
defined through the relation [see Eq.\ (\ref{eq:Qabcd})]
\begin{equation}
Q^{a c}(k) \, \equiv \, Q^{abcb}_{\mu \mu}(k) \, \equiv \,
\delta^{a c} N_c \, P_{\mu \mu}(k) \, Q(k^2) \, ,
\end{equation}
are shown in Figs.\ \ref{figs1} and \ref{figs2} as a function of $p^2(k)$. [We
indicate with $p(k)$ the lattice momentum with components $p_{\mu} = 2
\sin( \pi k_{\mu} / N)$, where $N$ is the lattice side and $k$ is the
wave vector with components $k_{\mu} = 0, 1, \ldots, N-1$.] Note that, in the latter
case, we plot the scalar function $Q(k^2)$ multiplied by $p^4$, in order
to make evident the IR behavior of the Bose-ghost propagator.
The data scale quite well, even though small deviations are observable in
the IR limit (see Fig.\ \ref{figs2}).

We also fit the data using the fitting function
\begin{equation}
\label{eq:fit}
f(p^2) \, = \, \frac{c}{p^4} \, \frac{p^2 + s}{p^4 \, + \,
                           u^2 p^2 \, + \, t^2 } \, .
\end{equation}
Following the analysis in \cite{Zwanziger:2009je,Zwanziger:2010iz},
this fitting function corresponds to considering an infrared-free
(Faddeev-Popov) ghost propagator $G(p^2)$ and a massive gluon propagator $D(p^2)$.
The fit describes the data quite well. As a consequence, the Bose-ghost
propagator presents a $p^{-4}$ singularity in the IR limit.
This result is in agreement with the one-loop analysis carried out in
\cite{Gracey:2009mj}. One should stress that, even though a double-pole
singularity is suggestive of a long-range interaction, the above result
does not imply a linearly-rising potential between quarks
\cite{Zwanziger:2009je,Zwanziger:2010iz,Gracey:2009mj}. Indeed, when
coupled to quarks via the $A-\phi$ propagator ---which is nonzero due
to the vertex term $\overline{\phi}_{\mu}^{ac} \, g f^{acb} A_{\nu}^{c}
\p_{\nu} \, \phi^{bc}_{\mu}$ in Eq.\ (\ref{saux})---, the Bose-ghost propagator
gets a momentum factor at each vertex \cite{Zwanziger:2009je,Zwanziger:2010iz},
i.e.\ the effective propagator is given by $p^{-2}$ in the IR limit.

\section{Conclusions}

We presented the first numerical evaluation of the
Bose-ghost propagator in minimal Landau gauge. We find
that our data are well described by a simple fitting function, which
can be related to a massive gluon propagator in combination with an
IR-free (Faddeev-Popov) ghost propagator, implying
a $p^{-4}$ singularity in the IR limit.
Our results constitute the first numerical manifestation of
BRST-symmetry breaking due to the restriction of the functional
integration to the Gribov region $\Omega$ in the GZ approach.
This directly affects continuum functional studies in
Landau gauge, which usually employ lattice results
as an input and/or as a comparison. At the same time,
several questions are still open for
a clear understanding of the GZ approach.
In particular, one should understand how a physical positive-definite
Hilbert space could be defined in this case.


\end{document}